# Measuring phonon dispersion at an interface

Ruishi Qi[1,2#], Ruochen Shi[1,2#], Yuanwei Sun[1,2], Yuehui Li[1,2], Mei Wu[1,2], Ning Li[1,2,3], Jinlong Du[1], Kaihui Liu[4,5,6], Chunlin Chen[7], Dapeng Yu[8], En-Ge Wang[2,9,10], & Peng Gao[1,2,5,6*]

**The breakdown of translational symmetry at heterointerfaces leads to the emergence of new phonon modes localized near the interface[1]. These interface phonons play an essential role in thermal/electrical transport properties in devices especially in miniature ones wherein the interface may dominate the entire response of the device[2]. Knowledge of phonon dispersion at interfaces is therefore highly desirable for device design and optimization. Although theoretical work has begun decades ago[1,3-6], experimental research is totally absent due to challenges in achieving combined spatial, momentum and spectral resolutions required to probe localized phonon modes. Here we use electron energy loss spectroscopy in an electron microscope to directly measure both the local phonon density of states and the interface phonon dispersion relation for an epitaxial cBN-diamond heterointerface. In addition to bulk phonon modes, we observe acoustic and optical phonon modes localized at the interface, and modes isolated away from the interface. These features only appear within ~ 1 nm around the interface. The experimental results can be nicely reproduced by *ab initio* calculations. Our findings provide insights into lattice dynamics at heterointerfaces and should be practically useful in thermal/electrical engineering.**

1 Electron Microscopy Laboratory, School of Physics, Peking University, Beijing 100871, China.
2 International Center for Quantum Materials, Peking University, Beijing 100871, China.
3 Academy for Advanced Interdisciplinary Studies, Peking University, Beijing 100871, China.
4 State Key Laboratory for Mesoscopic Physics, School of Physics, Peking University, Beijing 100871, China.
5 Collaborative Innovation Center of Quantum Matter, Beijing 100871, China.
6 Interdisciplinary Institute of Light-Element Quantum Materials and Research Center for Light-Element Advanced Materials, Peking University, Beijing 100871, China.
7 Institute of Metal Research, Chinese Academy of Sciences, Shenyang 110016, China.
8 Shenzhen Key Laboratory of Quantum Science and Engineering, Shenzhen 518055, China.
9 Songshan Lake Materials Lab, Institute of Physics, Chinese Academy of Sciences, Guangdong 523808, China
10 School of Physics, Liaoning University, Shenyang 110136, China
# These authors contribute equally to this work.
* Corresponding author. E-mail: p-gao@pku.edu.cn.

Understanding vibrational properties of condensed matter systems is important because phonons are involved in many physical phenomena ranging from thermal transport and superconductivity to mechanical strength and phase transition[1,7]. At a heterointerface, the presence of a boundary breaks the translational symmetry and leads to localized phonon modes[1,8], which are essential in understanding interface thermal/electrical transport properties, interface structural stability, and interface superconductivity[8-12]. On the practical side, the development of various modern technologies, such as computing circuits, light-emitting diodes, quantum cascade lasers, and thermoelectrics, demands efficient thermal engineering of interfaces[2]. In opposition to the extensively used assumption that interface thermal conductance can be calculated by considering bulk phonon modes in two constituting materials only, recent theoretical works have demonstrated that a large additional contribution from localized interfacial modes must be considered[2,8,10,13]. It is thus highly desirable to measure the dispersion relation of these localized modes, which directly dictates the transport properties across the interface.

Despite their great importance, related studies, especially experimental ones, are deficient. Interfacial localized modes have not been directly observed by spatial-resolved experiments to the best of our knowledge, not to mention measuring their dispersion relation. The major reason is that localized modes decay exponentially away from the interface with a characteristic spatial scale of only several atomic spacings, imposing stringent requirements to probe them experimentally. Sub-nanometer spatial resolution, milli-electronvolt spectral resolution, large momentum transfer and ultra-high sensitivity must be achieved at the same time. Inelastic neutron/X-ray scattering techniques have been successfully employed to study lattice dynamics of bulk crystals, but their beam size and detection sensitivity are orders of magnitude away from our goal[14]. Tip-enhanced Raman spectroscopy and infrared spectroscopy can recently achieve nanometer spatial resolution[15,16], but the tiny momentum transfer of photons precludes the possibility to gain information in the whole Brillouin zone (BZ). Atom-surface scattering spectroscopy and high-resolution reflection electron energy loss spectroscopy (EELS) are able to determine the full surface phonon dispersion relations[17], but their non-penetrating geometry makes interface measurements impossible. Without the stimulation and verification from experimental studies, the progress in related theoretical and computational works has been slow.

With recent advances in scanning transmission electron microscopes (STEMs), nanoscale vibrational STEM-EELS measurements have become possible in the past decade. This technique offers electron probes with high spatial, momentum and energy resolutions in conjunction with a large momentum transfer range and high detection sensitivity[18-20]. It has been successfully employed in many nanoscale phonon measurements[18,21-30], and in particular, in measuring phonon density of states (DOS) of single defects[26,27], and phonon dispersion of graphene sheets[28] and BN crystals[20]. Despite the progress, due to the intrinsic trade-off between the spatial and momentum resolutions previous studies either achieved atomic spatial resolution upon losing momentum resolution[25,26,31] or achieved fine momentum resolution upon sacrificing the beam size to several tens of nanometers[20,28,32]. Requiring both resolutions at the same time, the measurement of local interface phonon dispersion has remained elusive.

In this study, we use the recently developed four-dimensional EELS (4D-EELS) technique[22] to probe the interface phonon modes of cubic boron nitride (cBN) - diamond heterointerface. After carefully balancing the spatial and momentum resolutions, we successfully extracted the local DOS and the interface phonon dispersion curve. We observe both acoustic and optical branches of localized modes, and in addition, a branch of isolated modes with reduced vibration at the interface. All these features are confined within ~ 1 nm around the interface. The experimental results can be nicely reproduced by density functional perturbation theory (DFPT) calculations. Our work also demonstrates the possibility to measure the full phonon dispersion curve, not merely the DOS, of nanostructures such as interfaces and crystal defects, providing new opportunities in condensed matter physics, solid-state chemistry, material science and engineering.

As a versatile technique, STEM-EELS can collect vibrational signal in various configurations. In general, a kiloelectronvolt electron beam is focused on the sample, launching lattice vibrations upon losing a tiny fraction of its kinetic energy. An EELS aperture on the diffraction plane is used to selectively collect scattered electrons with particular momentum transfers. The collected electrons pass through a spectrometer and finally produce an EEL spectrum. To probe local lattice dynamics, sub-nanometer spatial resolution and fine momentum resolution are both desired. We first measure the local phonon DOS by choosing a large beam convergence angle to achieve a spatial resolution down to atomic scale (Fig. 1a). A large round EELS aperture collects electrons with a wide range of momentum transfers, producing EEL spectra containing information of phonon DOS. With the beam scanning in two spatial dimensions ($x$ and $y$), a three-dimensional EELS (3D-EELS; $x$-$y$-$\omega$, where $\omega$ is the frequency) dataset can be recorded. This allows us to map the spatial distribution of interfacial modes precisely. To ultimately obtain the dispersion curve, we then use another configuration (illustrated in Fig. 1b) to achieve the best balance between spatial and momentum resolutions. With a medium convergence angle, the spatial resolution is sacrificed but still at nanometer level (~ 1.5 nm in our experiment), and meanwhile a reasonable momentum resolution (~ 1/4 BZ size) can be achieved. With a slot EELS aperture, the dispersion diagram along a selected momentum line can be recorded in parallel, yielding a 4D-EELS ($x$-$y$-$\omega$-$q$, where $q$ denotes momentum transfer) dataset as the beam scans on the sample[22].

We choose high-quality epitaxial cBN-diamond heterojunction as a model system[33]. Diamond and cBN are top two thermal-conducting materials[34,35], making them promising in industrial applications; they can form atomically flat interfaces in large regions that facilitates nanoscale phonon studies[36]. Fig. 1c-d show two integrated differential phase contrast (iDPC)-STEM images of the interface viewed from two representative zone axes $[11\bar{2}]$ and $[1\bar{1}0]$, and Fig. 1e-f are corresponding projections of the atomic model for the interface structure. The same sample has also been carefully characterized elsewhere[36], confirming the atomic abruptness of the interface. At the interface, cBN is terminated with boron atoms that are directly bonded to carbon atoms on the diamond side[36].

Fig. 2a shows bulk phonon dispersion curves of cBN and diamond, as calculated by DFPT. Owing to similar crystal structure and atomic masses, their phonon dispersions are similar, with the exception of the TO branch. As a polar material, cBN has a finite LO-TO splitting gap due to long-range Coulomb interaction, which is absent in diamond. Thus, the DOS peak of cBN TO phonon is lower than that of diamond, while their acoustic DOS peaks have a large overlap region. Now we focus on the interface between them. It is well-known that surface phonon modes lying below bulk bands will appear at crystal surfaces[1,37,38]. At interfaces, however, phonon modes are even more complicated and related theories have yet to mature. Recent theoretical works classified phonon modes near interfaces into four categories[8,13]: extended modes that involve vibrations on both sides, partially extended modes that involve vibrations only on one side, interfacial modes that are localized at the interface, and isolated modes that are isolated away from the interface. To verify these predictions, we acquired 3D-EELS datasets in a clean interface region with no nearby defects observed. Fig. 2b shows an EELS line profile across the interface, with the beam travelling along the $[11\bar{2}]$ direction. The phonon signal changes abruptly near the interface within a spatial distance of ~ 1 nm (a few atom layers). Meanwhile, the spectrum near the interface is not simply a linear combination of two bulk spectra. For the cBN side, the LO phonon significantly reduces its intensity upon approaching the interface (blue arrow); the TO mode and acoustic modes show less new features near the interface, only with a tiny redshift. For the diamond side, the optical phonon peak has a large redshift at the interface (white arrow), while the acoustic peak has less features. Extended Data Fig. 1 further proves the interface spectra cannot be obtained by linear combinations of two bulk spectra. Even if we perform a least-square fitting to find the linear combination that matches the acquired spectra best, near the interface there is still a large residual that originates from new modes localized at the interface. We acquired such data for multiple times in different regions, and got highly reproducible results (Extended Data Fig. 2).

To understand and corroborate our experimental results, we performed DFPT phonon calculations on a (111) interface model. The calculated projected DOS (PDOS) of each atom layer is shown in Fig. 2c. For the carbon layers near the interface, the projected DOS is considerably different from that of the bulk diamond. The energy of the optical phonon peak is about 10 meV lower than bulk diamond, and this effect is localized within about three carbon layers. For cBN layers near the interface, qualitatively similar but less prominent redshift behavior is also observed for the TO peak, and the LO peak almost vanishes at the interface. We further calculated the phonon scattering cross section by approximating the electron probe as a Gaussian beam, where the finite beam size, experimental energy resolution, and EELS collection geometry are all considered (Methods). The calculated EELS line profile is displayed in Fig. 2d, which agrees nicely with the experiment. The redshift of diamond optical phonon and the intensity decay of cBN LO phonon are both well reproduced by the calculation. For a more quantitative comparison, Fig. 2e and Fig. 2f show measured and calculated spectra with the beam located in cBN (green curve), in diamond (blue curve), and at the interface (red curve). The interface component that cannot be obtained from bulk spectra, as extracted by the aforementioned fitting residual (Extended Data Fig. 1), is plotted as the black curves. This residual represents the DOS of the interface phonon modes excluding bulk contributions. By comparing the relative intensity among spectra acquired in different positions, all four types of modes (extended, partially extended, interfacial, and isolated) can be recognized in different energy regions.

For the overlapping region of two bulk acoustic DOS peaks, no remarkable new features were observed at the interface, so the intensity map (top panel of Fig. 2g) is almost homogenous just as the interface is transparent. This corresponds to an extended phonon mode that involves vibrations at both sides. The optical DOS peaks of the two materials have little overlapping, so the intensity maps for the corresponding energy windows (second and third panels in Fig. 2g) show an abrupt intensity change at the interface, characteristic for partially extended modes that involve vibrations predominately on one side. Interestingly, we identified three energy regions in which the interface spectrum is higher than both sides. We interpret them as localized interfacial modes. In the fourth to sixth panels in Fig. 2g, the vibration signal shows a substantial enhancement at the interface, decaying rapidly away from the interface. From these three maps, the root-mean-square width of the enhancement is found to be 1.4 nm, 1.0 nm and 0.8 nm, respectively, in accordance with the degree of localization of the calculated PDOS. The first mode (60-74 meV) is less localized mainly due to the existence of bulk acoustic modes at the same energy, forming an interface resonance mode as a mixture of the interfacial localized mode and the bulk continuum. For the other two modes, the bulk DOS at the same energy is small, so their localization is within 1 nm. By collecting scattered electrons at high-order BZs (off-axis geometry, Extended Data Fig. 3), one of them can be visualized with a higher contrast, but the main spectrum features are consistent. In addition, at ~ 160 meV the interface spectrum is considerably weaker than both sides (bottom panel in Fig. 2g). This could be an isolated mode that has significantly reduced vibration amplitude at the interface. These observations directly prove the existence of the interface phonon mode, and verify previous theoretical predictions.

Thus far we have used a large convergence angle to measure local DOS at near atomic resolution, but the collected signal is averaged over the whole BZ. However, interface thermal conductance and many other properties depend not only on the DOS, but also on the phonon group velocity and dispersion relation that cannot be obtained without momentum resolution[2]. To gain insights into the dispersion curve in the momentum space, we next employ 4D-EELS technique to directly measure the phonon dispersion of the interfacial modes. At the interface, the phonon dispersion can only be defined in a two-dimensional interface BZ (a slice of the bulk BZ) due to the broken translational symmetry along the third dimension (Extended Data Fig. 4). We place the slot aperture along ΓΣKX line, the common high-symmetry line of the bulk BZ and the interface BZ (inset in Fig. 1a). Although the dispersion can be measured nicely with smaller convergence angles (Extended Data Fig. 4), the required nanometer spatial resolution leads to a better balance at a medium angle (Methods). Fig. 3a shows measured dispersion diagrams with the beam located in cBN, at the interface, and in diamond, and Fig. 3b is corresponding DFPT simulation results. For completeness, EELS line profiles at five momentum transfers are given in Extended Data Fig. 5. Although the interface spectra seem to be simply an average of two bulk spectra, subtle difference between them can be recognized by subtracting the averaged bulk spectra from the interface spectra (i.e., interface – cBN/2 – diamond/2), as shown in Fig. 3c-d. The experimental diagram in Fig. 3c is sensitive to noise after this subtraction, but decent agreement with the calculation is still achieved. Interfacial localized modes appear below 150 meV, where both acoustic and optical branches are recognizable. The acoustic branch has increasing energy from BZ center to BZ boundary, while the optical branch is largely dispersionless. The relatively large group velocity of the acoustic interfacial mode is expected to give a large contribution to the thermal conductance. In addition, in Fig. 3c-d there is a dispersion line with negative intensity at 150-160 meV, which reflects a reduced vibration amplitude at the interface. The experimental results here can be validated not only by the calculation, but also by its consistency with results shown in Fig. 2. The flat interfacial optical band gives rise to an enhanced phonon PDOS at ~135 meV, and the negative signal here sums up to a decreased signal at ~160 meV. This cross-validates the high-resolution phonon maps in Fig. 2.

To better interpret these spectra, the calculated phonon dispersion of the interface model projected onto the (111) interface is shown in Fig. 3e, where three color channels are used to represent vibration amplitude at the interface (red), in cBN (green) and in diamond (blue). Green and blue regions originate from projected bulk phonon bands (Extended Data Fig. 6). Apart from them, we observe new dispersion lines (reddish lines) localized at the interface. The fact that interface modes form well-defined narrow dispersion lines further confirms they are new phonon modes highly localized at the interface, not from a gradual change of bulk modes. For three representative modes marked by arrows in Fig. 3d, Fig. 3f plots their calculated phonon eigenvectors. For both interfacial acoustic and optical modes (top and middle panels), the vibration amplitude is large at the interface, decaying to nearly zero within a few atom layers; the isolated mode (bottom panel) has considerable vibration amplitudes in both materials away from the interface, but has a vanishingly small amplitude at the interface. Thus, the measured positive signal in Fig. 3c directly reveals the dispersion relation of the localized interfacial modes, and the negative line gives the dispersion relation of the isolated mode.

In summary, we have visualized interface phonon modes using STEM-EELS. Atomic-scale phonon maps reveal modes localized at the interface and modes isolated away from the interface, whose features only appear within ~ 1 nm around the interface. 4D-EELS measurements further reveal their dispersion relation. Such measurements should be useful for understanding interface properties as needed by both fundamental researches and industrial applications. This work also demonstrates the possibility to perform phonon dispersion measurements on other important interfacial systems, such as topological phononic materials[39-41] and interface superconductors[12].

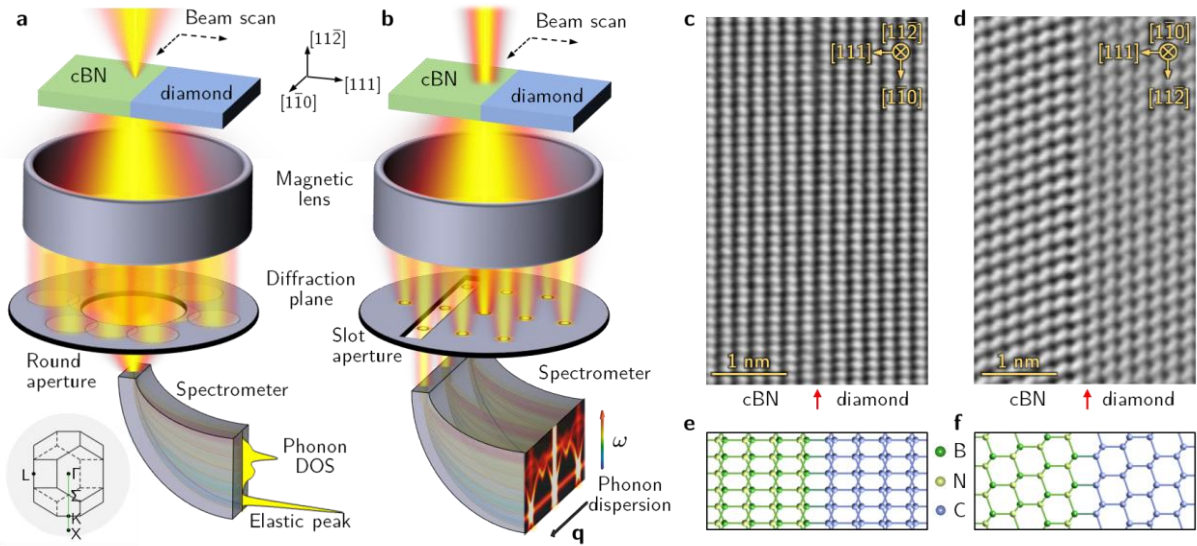

**Fig. 1 | Experimental setup and interface structure. a**, Schematic of the experimental setup (3D-EELS) used to acquire phonon DOS at atomic resolution. Inset at the bottom-left corner, top view of the bulk BZ with high-symmetry points marked. The two-dimensional interface BZ, viewed from this direction, is the vertical central line. **b**, Schematic of the experimental setup (4D-EELS) used to acquire phonon dispersion curves. **c-d**, iDPC-STEM images of the coherent cBN-diamond interface, viewed from the $[11\bar{2}]$ zone axis (**c**) and the $[1\bar{1}0]$ zone axis (**d**). Red arrows mark the interface. **e-f**, Corresponding projections of the atomic model.

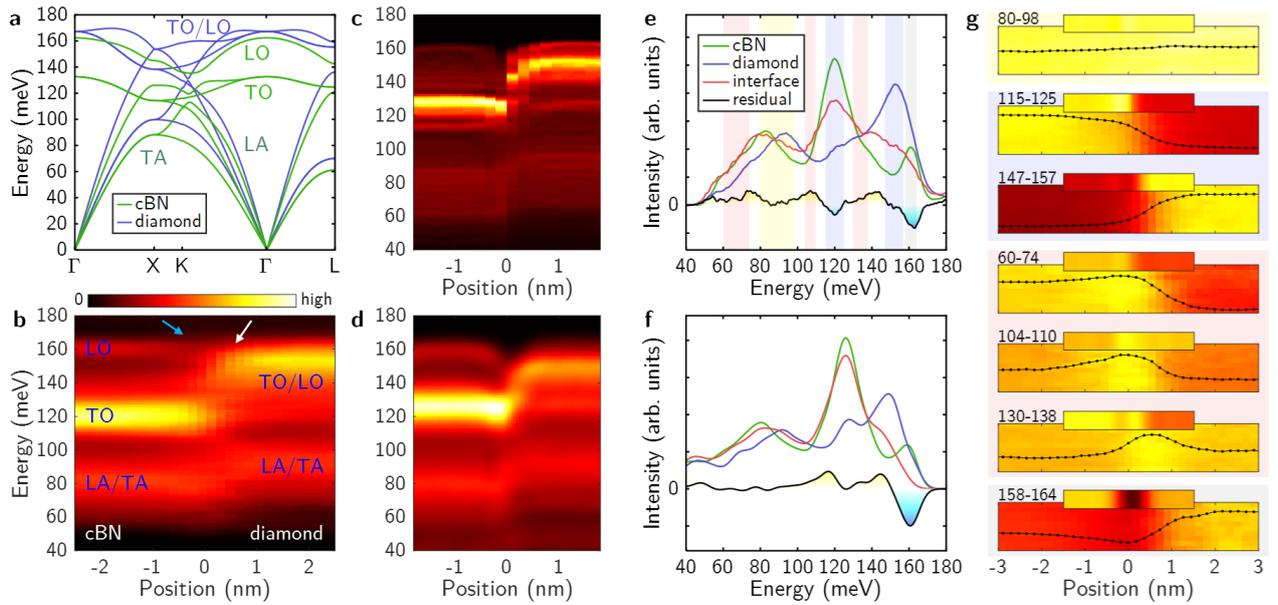

**Fig. 2 | Interface phonon measurement at atomic scale. a**, Calculated phonon dispersion for bulk cBN (green) and diamond (blue). **b**, Measured EELS line profile across the interface. The origin of the horizontal axis is at the interface. Blue and white arrows are guides to the eye that mark the intensity decrease of cBN LO phonon and the energy shift of the diamond optical phonon. **c**, Calculated phonon DOS projected onto atom layers. **d**, Calculated EELS scattering cross section. **e**, EEL spectra acquired in cBN (green), in diamond (blue), and at the interface (red). The black curve with gradient filling is the interface component that cannot be expressed as a linear combination of two bulk spectra. **f**, Corresponding calculation results. **g**, Energy-filtered EELS maps. Overlaid lines are averaged intensity profiles, whose contrast has been adjusted for clarity. Small inset bars at the top of each map are calculated maps. Energy integration windows are indicated by numbers (in meV) at the top-left corner of each map, and are also shaded in **e**. In all colormaps, the lower color limit is fixed to zero.

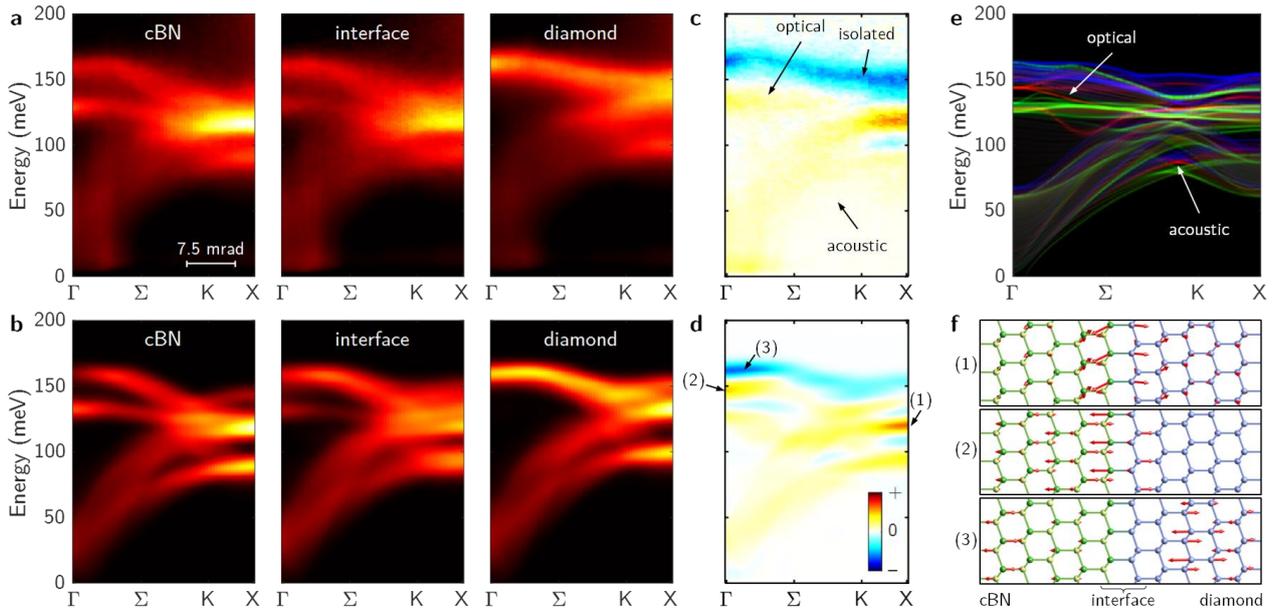

**Fig. 3 | Phonon dispersion measurements at the interface. a**, Measured phonon dispersion diagrams along ΓΣKX line with beam located in cBN (left), at the interface (middle) and in diamond (right). **b**, Corresponding simulation results based on DFPT calculation. **c-d**, Measured (**c**) and calculated (**d**) difference between the interface spectra and the average of two bulk spectra (i.e., interface – cBN/2 – diamond/2). **e**, Calculated phonon dispersion of the interface model, projected onto the (111) interface. Three RGB channels represent the squared norm of the vibration eigenvector of the interface layers (red), cBN (green) and diamond (blue). **f**, Phonon eigenvectors for three typical modes (with their momentum and energy labeled in **d**). Red arrows are the real part of eigenvectors showing the vibration amplitude of each atom.

# Methods

**Sample preparation.** The cBN crystals were grown epitaxially on the diamond substrates by the temperature gradient method under high temperature and high pressure conditions[33]. TEM samples were prepared by focused ion beam (FIB) technique (Hitachi FB2200 FIB) with a gallium ion source. To clean the damage layer of TEM samples induced by the ion radiation of FIB, Ar ion-milling with an accelerating voltage of 0.5 kV was performed using a precision ion polishing system (Model 691, Gatan).

**EELS data acquisition.** The EELS data was acquired on a Nion U-HERMES200 microscope equipped with both a monochromator and aberration correctors. 3D-EELS datasets were acquired with 60 keV beam energy, 35 convergence semi-angle and 25 mrad collection semi-angle. For datasets labeled "off-axis", the central diffraction spot was fully displaced away from the EELS aperture; for other datasets no such displacement was applied. Typical dwell time was 400 – 800 ms per pixel and ~ 30 min for each dataset in total. Sample drift is usually within 1 nm on this time scale, and is corrected afterward by aligning the interface. Typical energy resolution (full width at half maximum of the elastic line) under these conditions is 10 – 12 meV. Typical spatial resolution is 0.2 nm.

4D-EELS datasets were acquired with 30 keV beam energy and 7.5 convergence semi-angle. A slot aperture with aspect ratio 16:1 was placed along the $\Gamma K X K \Gamma$ line. To avoid the central diffraction spot and enhance the signal-to-background ratio, the aperture was displaced along $\Gamma L$ direction by a reciprocal lattice vector. Typical dwell time was 15 – 20 s per pixel and ~ 40 min for each dataset in total. Typical energy resolution is 10 – 12 meV. Typical spatial resolution is 1.5 nm.

**EELS data processing.** All acquired spectra were processed by custom-written Matlab code. For each dataset, EEL spectra were first registered by their normalized cross correlation to correct beam energy drifts. After the alignment, we applied block-matching and 3D filtering (BM3D) algorithm to remove Gaussian noise[42,43]. The data is individually denoised in two spatial dimensions for each energy and momentum channel, where the noise level is estimated based on high-frequency elements in the Fourier domain.

For 3D-EELS datasets, the spectra are normalized by the ZLP total intensity. The ZLP was removed by fitting the spectra to a Pearson function in two energy windows, one before and one after the energy loss region (approximately 20-45 meV and 180-220 meV, but slightly adjusted for each dataset to achieve the best fitting). Lucy-Richardson deconvolution was then employed to ameliorate the broadening effect caused by finite energy resolution, taking the elastic peak as the point spread function.

For 4D-EELS datasets, a correction for the statistical factor is performed following literature[44]. This process suppresses low-energy peaks because they have a higher occupation number. After the correction, the elastic peak is automatically vanishingly small and can be neglected.

**Ab initio calculations.** The DFPT calculations were performed within Quantum ESPRESSO[45,46] using Perdew-Zunger exchange-correlation functional and ultrasoft pseudopotential. This pseudopotential was chosen because it gives accurate bulk phonon frequencies. The interface model contains 12 layers of carbon atoms connected to 12 layers of cBN (48 atoms in one hexagonal unit cell with cell parameters $a$=2.518 Å and $c$=49.301 Å). The structure was optimized under $C_{3v}$ group symmetry constraint until the residual force is below $10^{-4}$ Ry/Bohr on every atom. The phonon dispersion and PDOS was calculated by interpolating the dynamical matrix on a $6 \times 6 \times 1$ mesh. Compared with a smaller model with 36 atoms in one unit cell, no noticeable change in phonon PDOS was observed.

For an infinite bulk crystal, the wavefunction of the electron beam can be treated as plane waves and then the scattering cross section can be calculated as[21,28,47]

$$\frac{d^2\sigma}{d\omega d\Omega} \propto \sum_{\text{mode }\lambda} |F_\lambda(\mathbf{q})|^2 \left[ \frac{n_q+1}{\omega_\lambda(\mathbf{q})} \delta(\omega - \omega_\lambda(\mathbf{q})) + \frac{n_q}{\omega_\lambda(\mathbf{q})} \delta(\omega + \omega_\lambda(\mathbf{q})) \right]$$

where $\omega_\lambda(\mathbf{q})$ and $n_q$ are the frequency and occupancy number of the $\lambda^{\text{th}}$ phonon mode with wavevector $\mathbf{q}$. The two terms in the square brackets correspond to phonon emission and absorption process respectively. The coupling factor

$$F_\lambda(\mathbf{q}) \propto \frac{1}{q^2} \sum_{\text{atom }k} \frac{1}{\sqrt{M_k}} e^{-i\mathbf{q}\cdot\mathbf{r}_k} e^{-W_k(\mathbf{q})} Z_k(\mathbf{q}) [\mathbf{e}_\lambda(k,\mathbf{q}) \cdot \mathbf{q}]$$

is determined by the mass $M_k$, real-space position $\mathbf{r}_k$, effective charge $Z_k(\mathbf{q})$, Debye-Waller factor[48] $\exp(-2W_k(\mathbf{q}))$ and phonon displacement vector $\mathbf{e}_\lambda(k,\mathbf{q})$ of $k^{\text{th}}$ atom in a unit cell. The effective charge $Z_k(\mathbf{q})$ was calculated following literature[47] with atomic form factors constructed from parameters in literature[49].

When the interface is concerned, approximating the beam as plane waves is no longer acceptable. We model it as a Gaussian beam. Then the (position-dependent) coupling factor can be approximated as

$$F_\lambda(\mathbf{q},\mathbf{R}) \propto \frac{1}{q^2} \sum_{\text{atom }k} \frac{1}{\sqrt{M_k}} G(|\bar{\mathbf{r}}_k - \mathbf{R}|) e^{-i\mathbf{q}\cdot\mathbf{r}_k} e^{-W_k(\mathbf{q})} Z_k(\mathbf{q}) [\mathbf{e}_\lambda(k,\mathbf{q}) \cdot \mathbf{q}]$$

where $\mathbf{R} = (X,Y)$ stands for the 2D position of the electron beam, the overbar on $\mathbf{r}_k$ means only taking its first two components, and $G$ is the Gaussian function with its width determined by the spatial resolution. The EEL spectra were simulated by summing in the momentum space in a region corresponding to the EELS aperture collection range, convoluted by the angle distribution of the beam (determined by the convergence semi-angle).

## Data availability

The data that support the findings of this study are available from the corresponding author upon request.

## Code availability

A GUI version of the matlab code for EELS data processing and DFPT related post-processing codes are available from the corresponding author upon request.

**Acknowledgements**

The work was supported by the National Key R&D Program of China (2019YFA0708200, 2016YFA0300903), the National Natural Science Foundation of China (11974023, 52021006, 12004010), Key-Area Research and Development Program of Guangdong Province (2018B030327001, 2018B010109009), the "2011 Program" from the Peking-Tsinghua-IOP Collaborative Innovation Center of Quantum Matter, Youth Innovation Promotion Association of CAS, and the Introduced Innovative R&D Team Project of "The Pearl River Talent Recruitment Program" of Guangdong Province (2019ZT08C321). We acknowledge Electron Microscopy Laboratory of Peking University for the use of electron microscopes. We acknowledge High-performance Computing Platform of Peking University for providing computational resources for the DFPT calculation. We thank Prof. Ji Feng, Mr. Qiangqiang Gu, Dr. Chenglong Shi and Dr. Tracy Lovejoy for helpful discussion. We thank Prof. Takashi Taniguchi in NIMS for providing the samples.

**Author contributions**

R.Q. and R.S. contributed equally to this work (order determined by a random process upon completion of the manuscript). P.G., R.S. and R.Q. conceived the project. C.C. prepared the TEM sample. R.S. designed and performed the EELS measurements. R.Q. wrote the data processing codes and analyzed the data. R.Q. and R.S. performed the DFPT and scattering cross section calculations under direction of E.-G.W. Y.S., M.W., and Y.L. acquired atomic-resolution iDPC images under direction of P.G. N.L., K.L. and D.Y helped the data analysis. R.Q., R.S. and P.G. finalized the manuscript. P.G. supervised the project. All authors contributed to this work through useful discussion and/or comments to the manuscript.

**Competing interests**

The authors declare no competing interests.

# Extended data figures

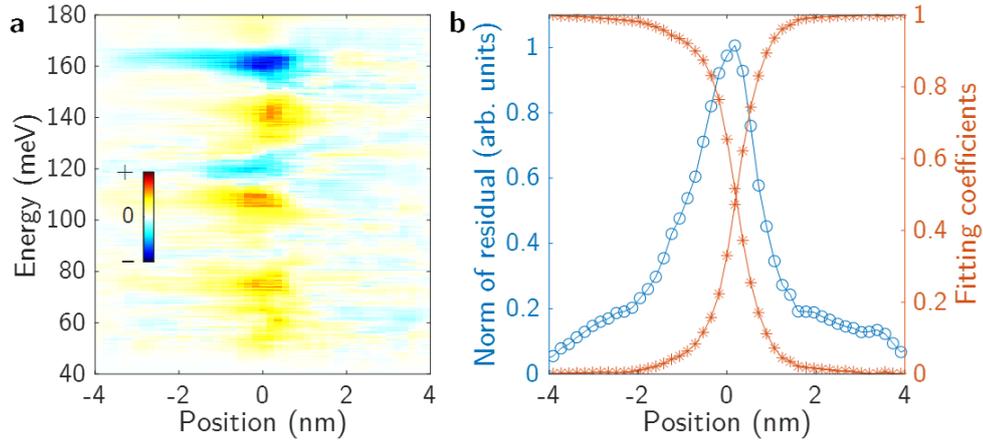

**Extended Data Fig. 1 | Interface component of the spectra extracted by finding the minimum difference between the measured spectrum and all possible linear combinations of two bulk spectra.** The fitting was performed by minimizing $\|S(\omega) - a_1 S_{\text{cBN}}(\omega) - a_2 S_{\text{diamond}}(\omega)\|$, where $S(\omega)$ is the measured spectrum (Fig. 2b), $S(\omega)$ with subscripts means the bulk spectra, and $a_1, a_2$ are adjusted coefficients. **a**, Line profile of the fitting residual. Since the fitting gives the linear combination that is closest to the measured spectrum, the residual represents the interface component that cannot be obtained from bulk modes. Near the interface, three red peaks correspond to three interfacial modes in Fig. 2g. The blue region at 160 meV is due to the isolated mode with reduced vibration at the interface. **b**, Norm (root sum squared) of residuals as a function of position (left axis), and the fitting coefficients (right axis). The residual is sharply peaked at the interface, indicating the interface mode is localized within ~ 1 nm near the interface.

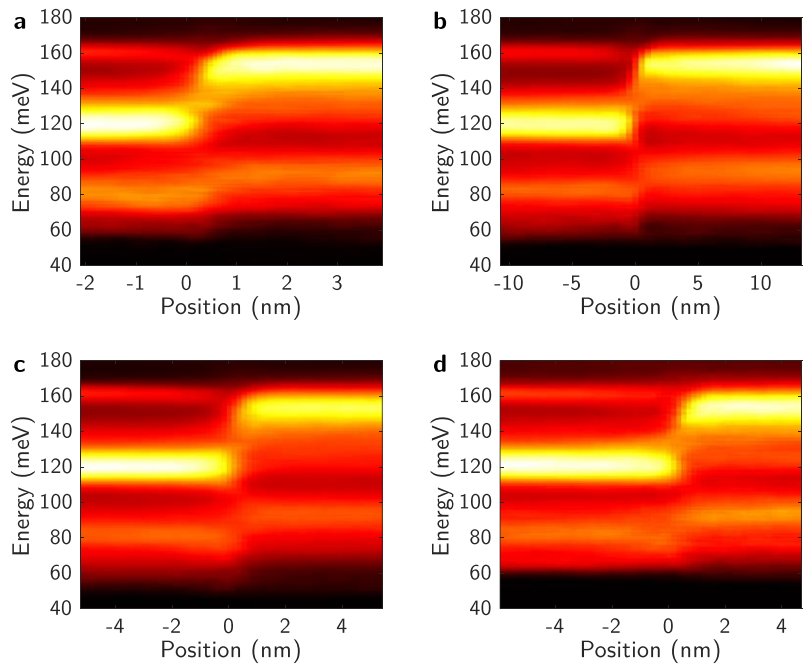

**Extended Data Fig. 2 | 3D-EELS data acquired in different regions. a-d**, Four EELS line profiles acquired under the same experimental conditions except different pixel sizes and different scanning regions. **a-b** and **c-d** were acquired in two experiments that are two weeks apart. All datasets give consistent results as the one shown in Fig. 2.

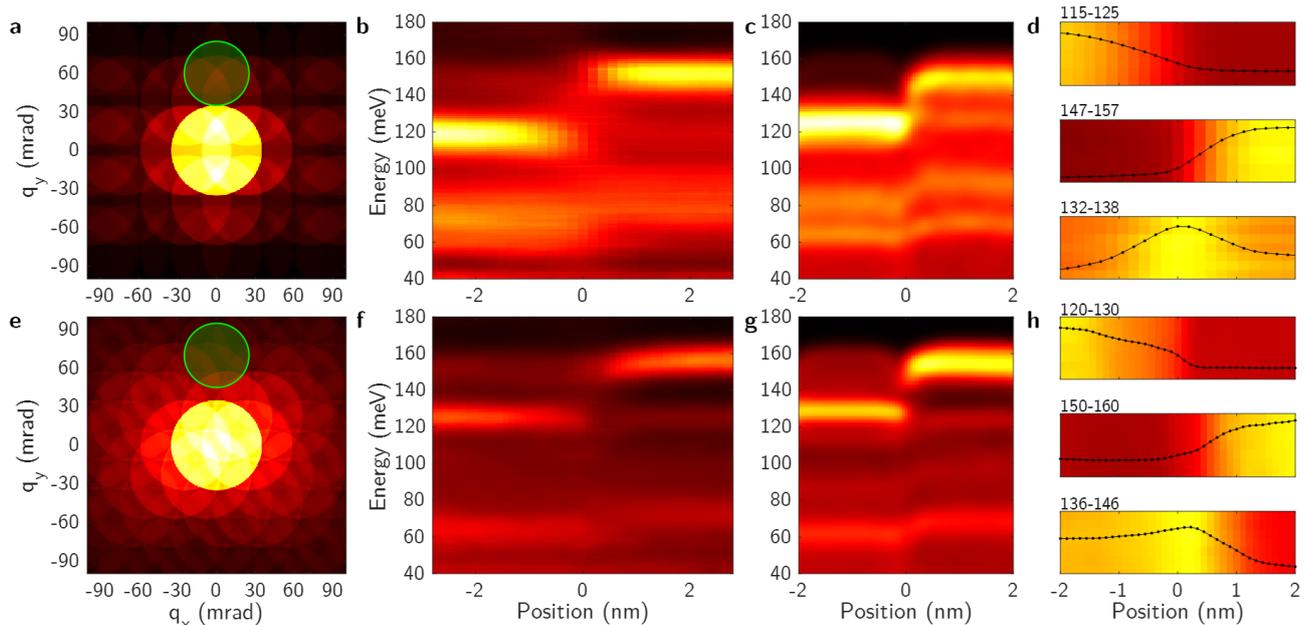

**Extended Data Fig. 3 | Off-axis EELS measurements. a**, Schematic of diffraction plane and EELS aperture placement. The colormap illustrates the diffraction plane viewed from [11$\bar{2}$] zone axis, with 60 keV beam energy and 35 mrad convergence semi-angle. The diffraction spot size (35 mrad) is larger than neighboring spot distance, so they partially overlap. The green circle marks the position of the slot aperture, which is displaced away from the central spot. **b**, EELS line profile acquired with off-axis geometry. Main spectral features are consistent with those acquired with on-axis geometry (Fig. 2b). **c**, Corresponding simulation result. **d**, EELS maps at selected energies. One of the interfacial modes has a better contrast than the on-axis result. **e-h**, same as **a-d**, but acquired along [1$\bar{1}$0] zone axis.

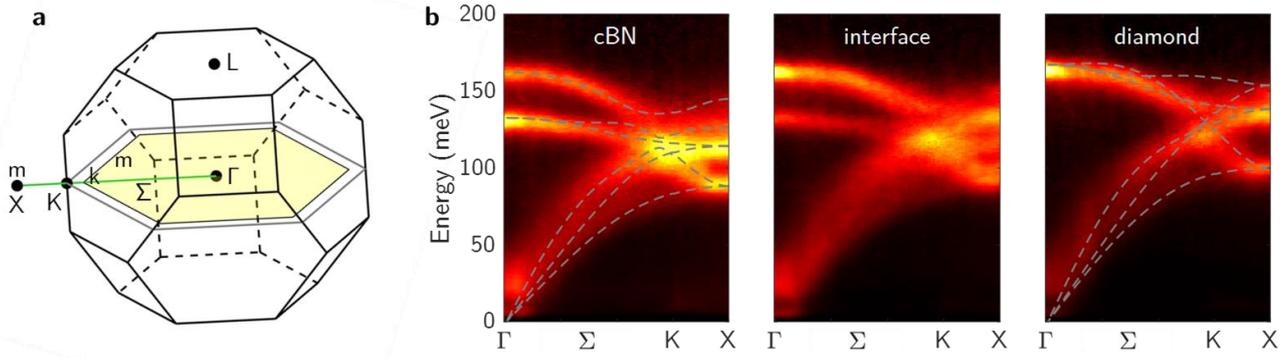

**Extended Data Fig. 4 | Phonon dispersion diagrams measured with 3 mrad convergence semi-angle. a,** Schematic of the bulk BZ (truncated octahedron) and the interface two-dimensional BZ (yellow hexagon). Upper-case and lower-case letters mark the high-symmetry points of the bulk BZ and interface BZ, respectively. **b**, Measured dispersion diagrams along ΓΣKX line with 3 mrad convergence semi-angle. Dashed curves are calculated bulk phonon dispersion. Although smaller convergence semi-angle gives better momentum resolution and hence nicer dispersion diagrams, insufficient spatial resolution makes it hard to extract localized features at the interface.

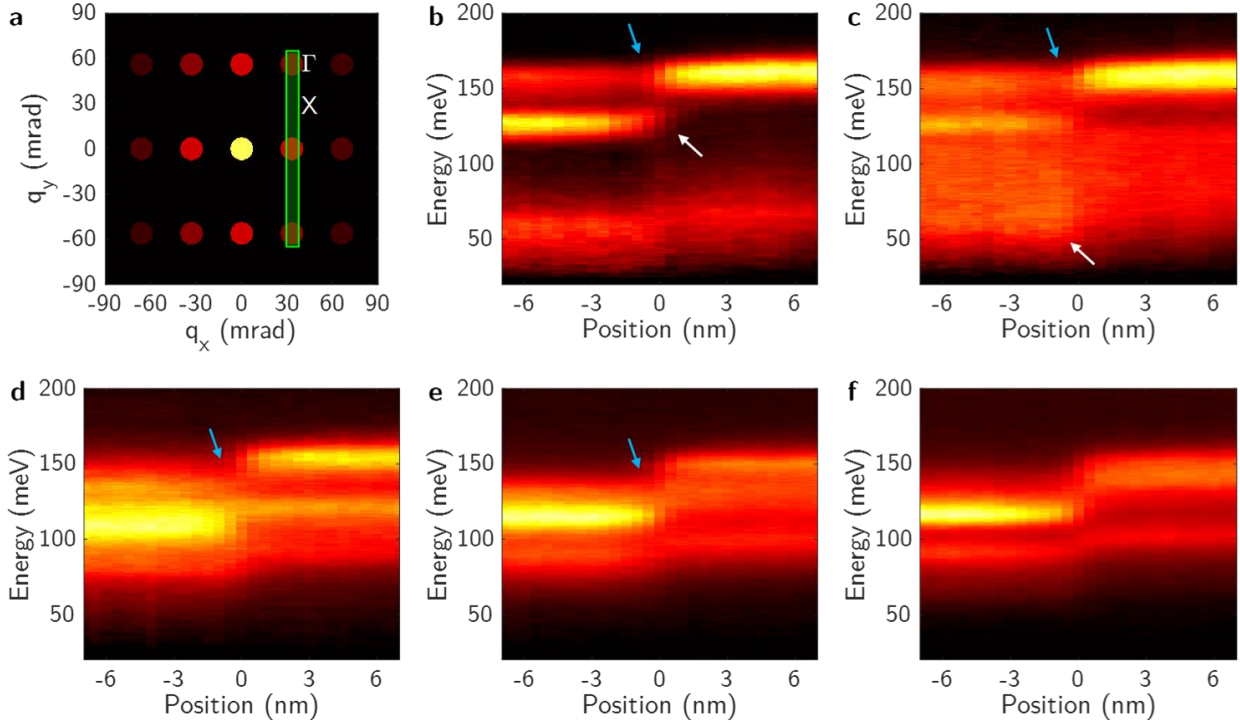

**Extended Data Fig. 5 | Spatial line profiles at five momentum transfers. a**, Schematic of diffraction plane and EELS aperture placement. The colormap illustrates the diffraction plane viewed from [11$\bar{2}$] zone axis, with 30 keV beam energy and 7.5 mrad convergence semi-angle. The diffraction spot size is drawn to scale, indicating our momentum resolution. The green rectangle marks the position of the slot aperture. **b-f**, Line profiles with momentum transfers from Γ (**b**) through the Σ line (**c-d**) to K (**e**) and finally X (**f**). The intensity decrease of the highest-frequency optical phonon is observable in most panels (blue arrows), which corresponds to the negative-intensity line at 150-160 meV in Fig. 3c-d. The interfacial mode is directly observable in some panels (white arrows).

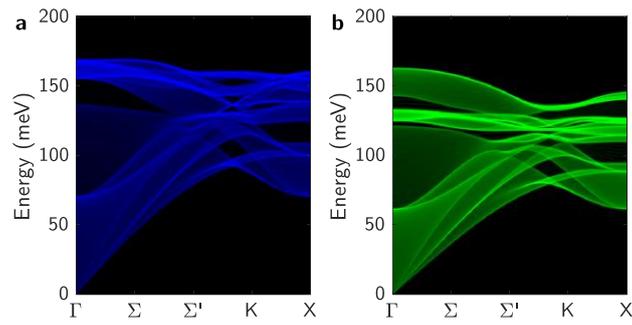

**Extended Data Fig. 6 | Projected bulk phonon bands. a-b**, Bulk phonon band of diamond and cBN projected onto (111) surface.